\documentclass[11pt,twoside]{article}

\usepackage{asp2006}
\usepackage{epsf}
\usepackage{lscape}

\begin{document}
\title{Modeling the Color Magnitude Relation for Galaxy Clusters}   
\author{N. Jim\'enez,$^{1,2}$ S. A. Cora,$^{1,2}$ A. Smith Castelli,$^{1,2}$ and L. P. Bassino,$^{1,2}$ }

\affil{$^1$Facultad de Ciencias Astron\'omicas y Geof\'isicas (FCAG) de la Universidad Nacional de La Plata (UNLP), Argentina.}

\affil{$^2$Instituto de Astrof\'isica de La Plata (CCT La Plata, CONICET)}

\begin{abstract}

We investigate the origin of the colour-magnitude relation (CMR) observed in 
cluster galaxies by using a combination of a cosmological {\em N}-body 
simulation of a 
cluster of galaxies and a semi-analytic model of galaxy formation. 
The departure of 
galaxies in the bright end of the CMR with respect to the trend denoted by 
less
luminous galaxies
could be explained by
the influence of minor mergers.
\end{abstract}

We consider a simulated galaxy cluster with virial mass 
$\approx 1.3 \times 10^{15} \, h^{-1} \, M_{\odot}$, 
\citep{Dolag05}. 
The semi-analtytic model of galaxy formation used \citep{Lagos08} 
considers gas cooling, star formation, galaxy mergers, disc instabilities, 
metal enrichment and feeback from supernovae and
active galactic nuclei.
We compare the photometric properties 
of simulated early-type galaxies
(those with a bulge component comprising $80$ per cent of the total 
stellar mass) 
with those of the  early-type galaxies in the central 
region of the Antlia cluster \citep{SmithCastelli08}, 
using the Washington photometric system. We find 
good agreement between the general trend of simulated and observed CMR. 
However, the more massive galaxies ($-22.<M_{\rm T_{\rm 1}}<-19.$) 
depart from the fit to 
observed data, displaying an almost constant colour 
($C-T_{\rm 1}\approx 1.7)$, 
as detected in other clusters. 

We select 
galaxies 
in six magnitude bins within the range $-22.<M_{\rm T_{\rm 1}}<-16.$
and  
analyse the evolution of their content of cold gas  
and the average number of minor mergers 
suffered by these galaxies. We find an increase of the latter quantity 
at low redshift for the more luminous galaxies. 
These galaxies are also characterized by a low content of cold gas which 
dramatically decreases since $ z \approx 2$. The combination of these two 
effects would yield to an increase of the luminosity without strongly 
affecting the galaxy colour. 
We will deepen the investigation on the origin of the CMR bright end 
with the aim of disentangling the role played by dry mergers in 
determining the properties of the more massive galaxies.

\end{document}